\PassOptionsToPackage{table}{xcolor}
\documentclass[sigconf]{acmart}
\usepackage{algorithm, algorithmic}
\usepackage{multirow}
\usepackage{hhline}
\usepackage{makecell}
\usepackage{balance}
\usepackage{enumitem}
\usepackage[table]{xcolor}
\AtBeginDocument{%
  }

\copyrightyear{2025} 
\acmYear{2025} 
\setcopyright{acmlicensed}\acmConference[MM '25]{Proceedings of the 33rd
 ACM International Conference on Multimedia}{October 27--31, 2025}{Dublin,
 Ireland}
 \acmBooktitle{Proceedings of the 33rd ACM International Conference on
 Multimedia (MM '25), October 27--31, 2025, Dublin, Ireland}
 \acmDOI{10.1145/3746027.3755705}
 \acmISBN{979-8-4007-2035-2/2025/10}

\begin{document}

\title{SpA2V: Harnessing Spatial Auditory Cues for Audio-driven Spatially-aware Video Generation}

\author{Kien T. Pham}
\affiliation{
  \institution{Hong Kong University of Science and Technology}
  \city{Clear Water Bay}
  \country{Hong Kong}
}
\email{tkpham@connect.ust.hk}

\author{Yingqing He}
\affiliation{
  \institution{Hong Kong University of Science and Technology}
  \city{Clear Water Bay}
  \country{Hong Kong}
}
\email{yhebm@connect.ust.hk}

\author{Yazhou Xing}
\affiliation{
  \institution{Hong Kong University of Science and Technology}
  \city{Clear Water Bay}
  \country{Hong Kong}
}
\email{yxingag@connect.ust.hk}

\author{Qifeng Chen}
\affiliation{
  \institution{Hong Kong University of Science and Technology}
  \city{Clear Water Bay}
  \country{Hong Kong}
}
\email{cqf@ust.hk}

\author{Long Chen}
\authornote{Long Chen is the corresponding author.}
\affiliation{
  \institution{Hong Kong University of Science and Technology}
  \city{Clear Water Bay}
  \country{Hong Kong}
}
\email{longchen@ust.hk}


\begin{abstract}
Audio-driven video generation aims to synthesize realistic videos that align with input audio recordings, akin to the human ability to visualize scenes from auditory input. However, existing approaches predominantly focus on exploring semantic information, such as the classes of sounding sources present in the audio, limiting their ability to generate videos with accurate content and spatial composition. In contrast, we humans can not only naturally identify the semantic categories of sounding sources but also determine their deeply encoded spatial attributes, including locations and movement directions. This useful information can be elucidated by considering specific spatial indicators derived from the inherent physical properties of sound, such as loudness or frequency. As prior methods largely ignore this factor, we present \textbf{SpA2V}, the first framework explicitly exploits these spatial auditory cues from audios to generate videos with high semantic and spatial correspondence. SpA2V decomposes the generation process into two stages: 1) \textit{Audio-guided Video Planning}: We meticulously adapt a state-of-the-art MLLM for a novel task of harnessing spatial and semantic cues from input audio to construct Video Scene Layouts (VSLs). This serves as an intermediate representation to bridge the gap between the audio and video modalities. 2) \textit{Layout-grounded Video Generation}: We develop an efficient and effective approach to seamlessly integrate VSLs as conditional guidance into pre-trained diffusion models, enabling VSL-grounded video generation in a training-free manner. Extensive experiments demonstrate that SpA2V excels in generating realistic videos with semantic and spatial alignment to the input audios.
\end{abstract}


\begin{teaserfigure}
  \includegraphics[width=\linewidth]{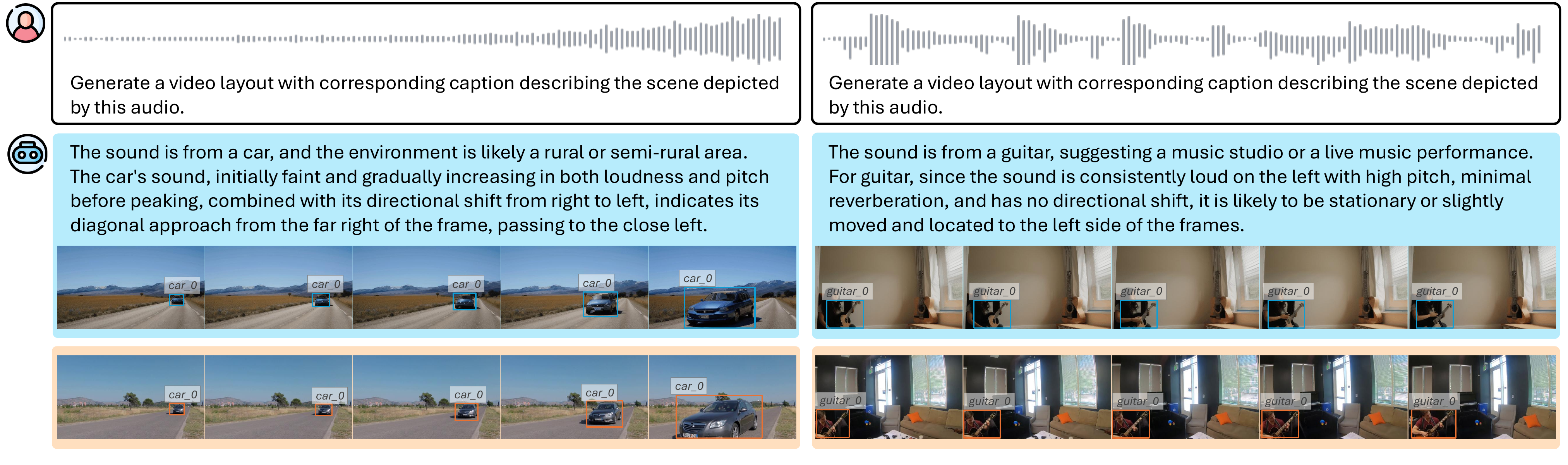}
  \caption{Audio-driven
Spatially-aware Video Generation targets to synthesize realistic videos that are semantically and spatially aligned with input audio recordings. Our proposed SpA2V framework accomplishes this task by decomposing generation process into two stages: \textit{Audio-guided Video Planning} and \textit{Layout-grounded Video Generation}, achieving audio-video correspondence via leveraging VSLs as intermediate representation to capture auditory cues and guide the generation process respectively. Here \colorbox{orange!25}{ground-truth videos} are for visual comparisons with \colorbox{cyan!25}{generated videos} only and are not inputted into our framework.}
  \Description{}
  \label{fig:teaser}
\end{teaserfigure}


\maketitle

\section{Introduction}
Content creation has witnessed a significant transformation in recent years, leading to a proliferation of novel creative tasks that were previously unimaginable. This evolution is driven by the emergence of various powerful generative models capable of generating and manipulating content in different modalities, including text~\cite{dou-etal-2022-gpt, gemini, llama3, gpt4, mm-llm}, image~\cite{sd3, sd1.5, gligen, tale, wang2025event, jiang2025clipdrag}, audio~\cite{audioldm, audioldm2, musicgen, tia2v}, and video~\cite{svd, ltx, cogvideox, animatediff, lian2024llmgrounded, gao2025ca2, videopoet}. Particularly in the context of video generation, the advancement is becoming more elusive with many current works making progress in synthesizing video content based on text description~\cite{cogvideox, ltx} and initial image~\cite{svd, cogvideox}. Despite showing impressive results, they often fall short in capturing the richness and temporal coherence of real-world events, because of the inherent ambiguity and static nature of their respective conditions. 

Audio, in contrast, naturally grounds video in reality and encodes abundant temporal and contextual information on sound-emitting objects, their interactions, and the spatial arrangement of the soundscape. These intrinsic values provide unique advantages for generating more nuanced, immersive, and temporally consistent video content, leading to more realistic and engaging experiences. In addition, similar to the human ability to use auditory information to depict corresponding visual scenes and events, audio-to-video generation can be applied to diverse applications that span across industry verticals. Some of these include automated scene visualization in filmmaking, dynamic product creation in multimedia, engaging advertisements in marketing, and accessible learning materials in education. In light of these significant advantages and useful applications, it is imperative to explore the field of \textbf{audio-driven video generation}. 

The prevailing audio-to-video generation methods typically rely on global semantic features extracted from audio tracks for synthesis. Although this approach can produce semantically aligned videos, it only works for specific simple soundscapes and often results in poor content quality and misaligned spatial composition with input audio in general scenarios. For example, existing works including~\cite{moda, saas, style2talk, eetalk} can generate talking head videos conditioned on speech, yet are not applicable to other domains. Other methods such as~\cite{chen2017, tempotoken, mmdiffusion, seeing-and-hearing} can synthesize videos of different contexts that are globally aligned with the specific semantic categories (e.g., dancing, drumming, landscape, etc.) of the input audio recording. However, their results lack spatial coherence between visual and auditory elements, affecting realism and ultimately diminishing the immersive experience. Some current approaches~\cite{sound2sight, linz2024asva} alleviate such an issue by directly providing an initial frame or video segment which already establishes a spatial correspondence with audio as an additional visual input, but inevitably limits the diversity of the generated content. 

Surprisingly, all the aforementioned works have largely overlooked the fact that sound inherently encompasses rich spatial information such as location and movement of sounding sources present in the scenes. Such information can be harnessed to generate according visual components with not only semantic but also spatial coherence to input audios. To this end, the first critical question that arises is: \textbf{Q1:} \textit{Can we directly decode the spatial information embedded within audio to drive video generation?}
\begin{figure}[t]
\includegraphics[width=\linewidth]{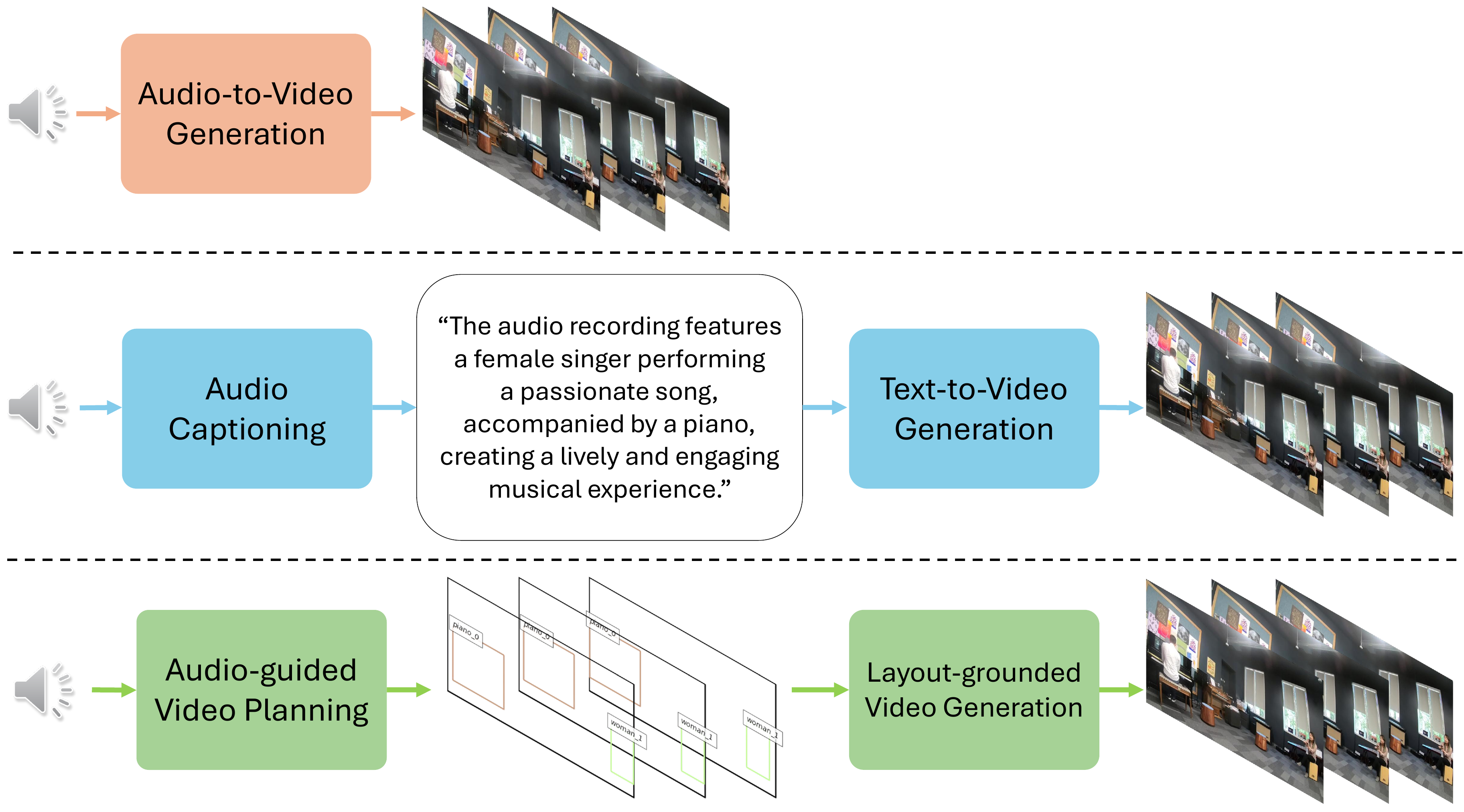}
  \caption{Different frameworks for audio-driven video generation. From top to bottom are the typical Audio $\rightarrow$ Video direct approach, two-stage Audio $\rightarrow$ Text $\rightarrow$ Video method, and our proposed novel Audio $\rightarrow$ Video Scene Layout $\rightarrow$ Video pipeline respectively.}
  \Description{}
  \label{fig:difference}
  \vspace{-1em}
\end{figure}
We draw inspiration from the fact that humans spontaneously perform similar tasks to perceive and navigate the environment in our daily hearing. We intuitively utilize our multisensory and commonsense knowledge to exploit specific auditory cues from environmental sounds, then reason on them to derive necessary information. For instance, considering the top-left example in Fig.~\ref{fig:teaser}, we can instinctively imagine an approaching car when hearing its engine sound getting louder. This is because we know what a car generally sounds and looks like (semantic clue) and deduce that increasing in volume (spatial clue) implies approaching motion. By targeting these auditory cues, we contemplate that a strong foundational model with human-like multimodal understanding and reasoning capabilities like MLLM has the potential to adapt and replicate this human instinct, driving us to explore it extensively to address this challenge. 

Once \textbf{Q1} is properly resolved, the important subsequent question that emerges is \textbf{Q2:} \textit{How should these information be represented to bridge the gap between audio and video modalities and guide the generation process?} At first thought, text description seems like a viable option. However, it suffers from inherent ambiguity, leading to inconsistent results and a lack of precise spatial control over scene composition in generation process. Video Scene Layout (VSL), on the other hand, offers a structured and unambiguous representation, enabling fine-grained manipulation of object placement and scene structure. Considering our concentration on spatial relationships between auditory and visual elements, VSL is intuitively advantageous compared to the textual counterpart. Therefore, we adopt it as our intermediate representation to capture the semantic and spatial attributes of the sounding sources extracted from input audio and then control the video generation process as shown in Fig.~\ref{fig:difference}.

We propose a novel framework dubbed \textbf{SpA2V} which is the first attempt to explicitly exploit spatial auditory information for video generation conditioning solely on audio. SpA2V decomposes the generation process into two respective stages, namely Audio-guided Video Planning and Layout-grounded Video Generation. The first stage is responsible for identifying sounding objects occurring in an input audio and inferring their semantic and spatial attributes to construct a VSL as guidance for generation in the subsequent stage. We employ a state-of-the-art Multimodal Large Language Model (MLLM), such as Gemini 2.0~\cite{gemini1.5} or GPT4o~\cite{gpt4o}, with demonstrated powerful understanding and reasoning capabilities across different modalities as the Video Planner for our SpA2V. We adapt them for our new task of audio-driven VSL generation via a meticulously designed prompting mechanism that leverages In-context Learning~\cite{incontext}, allowing it to effectively and efficiently harness semantic and spatial cues presented in input audio.

Following VSL generation, we synthesize the final video by conditioning on the VSL in the second stage. Our approach incorporates pre-trained diffusion models in an efficient and effective way, inspired by MIGC~\cite{migc} and AnimateDiff~\cite{animatediff}. These methods augment the pre-trained Stable Diffusion model with spatial grounding and motion modules for layout-to-image and text-to-video tasks. We exploit the fact that they train only these new modules while keeping the backbone intact. By directly integrating their learned modules into the same frozen backbone, we create a layout-to-video diffusion model capable of spatial grounding and motion modeling simultaneously without further training. We hereby employ it as our VSL-grounded video generator to complete this stage.  

To assess the capability of our SpA2V framework, we introduce a new benchmark named AVLBench curated from real-world stereo audio-video recordings~\cite{fairplay, vs13, vehicle, urbansas} and repurposed for our specific use cases. It includes diverse test scenarios featuring different numbers of sounding objects with various spatial attributes. Results from our experiments on this benchmark demonstrate that SpA2V achieves a high degree of semantic and spatial correspondence between the generated VSLs, videos, and the input audios, marking the first successful attempt of spatially-aware audio-to-video generation.   

Overall, our contributions are listed as follows:

\begin{itemize}[leftmargin=*]

    \item We propose a novel task of audio-driven spatially-aware video generation which aims at synthesizing videos with spatial correspondence to audio conditions.  
    \item We present SpA2V, the first framework attempting to fulfill the task by decomposing the generation process into two stages Audio-guided Video Planning and Layout-to-Video Generation and leveraging powerful pre-trained MLLMs and diffusion models to accomplish each stage, respectively.
    \item We introduce AVLBench, a new benchmark for evaluating alignment between input audios and generated VSLs and videos.
    \item Extensive experiments on the benchmark highlight the capability of SpA2V in generating realistic VSLs and videos where visual elements correspond both semantically and spatially to the sound sources in input audio. The implementation will be released on GitHub\footnote{\url{https://github.com/HKUST-LongGroup/SpA2V}}.
\end{itemize}
\section{Related Work}

\noindent\textbf{Audio-Visual Learning.}
Recent years have witnessed growing research efforts in audio-visual learning. Early studies primarily focused on cross-modal Audio-Visual Synchronisation~\cite{avs-wild,sparsesync,synchformer}, which employed self-supervised learning to align temporal relationships between audio and video, establishing foundational representations for downstream tasks. 
Despite resolving temporal alignment, these methods largely overlooked semantic and spatial correlations between audio and visual modalities.
To provide more detailed audio-visual spatial alignment and support the audio as a guiding signal, ~\cite{avs-semantics, CAVP-unraveling} explore the Audio-Visual Segmentation (AVS) task that pioneered the prediction of sounded object segmentation maps in video frames conditioned on audio input. However, they focused on perception rather than generation, limiting their applicability to generative tasks. Critically, they also neglected to model the spatial attributes of audio (e.g., sound source localization or motion trajectories), which are vital for grounding visual scenes in physical reality.
Recent advances have started to explore \textit{spatial audio} for audio-visual tasks. 
For example, BAT~\cite{zheng2024bat} leverages large language models (LLMs) for spatial sound reasoning, while ELSA~\cite{ELSA} learns spatially-aware language-audio representations for fine-grained localization. Building on these insights, our work is the first to explicitly exploit spatial audio cues for audio-guided video generation, enabling the synthesis of videos where visual elements are both semantically and spatially coherent with sound sources.

\noindent\textbf{Audio-to-Video (A2V) Generation.}
A2V generation focuses on producing visual content that aligns with given audio inputs. Several studies have explored this domain by leveraging audio to provide semantic cues and temporal dynamics for video generation.
Sound2Sight~\cite{sound2sight} and CCVS~\cite{le2021ccvs} utilize audio alongside preceding video frames to forecast subsequent frames, capturing visual dynamics driven by the input audio. ~\cite{soundguidevgen} employs StyleGAN, projecting audio into its latent space to navigate trajectories within this space, effectively aligning audio with visual content. 
Seeing and Hearing~\cite{seeing-and-hearing} introduces a diffusion latent aligner to synchronize audio with visual elements, enhancing the coherence between them. 
TempoTokens~\cite{tempotoken} adapts a pre-trained text-to-video diffusion model for A2V generation, aligning audio and visual components to improve synchronization.
Although these approaches focus on semantic and temporal alignment, they often overlook the spatial aspect when processing input audio. Spatial information such as the location and distance of sounded objects can bring significant enhancement to the generated results. In this work, we pioneer the exploration of harnessing important spatial cues from input audio to guide video generation and fulfill this gap. We term the task as \textbf{audio-driven spatially-aware video generation}.
\begin{figure*}
\includegraphics[width=1.0\textwidth]{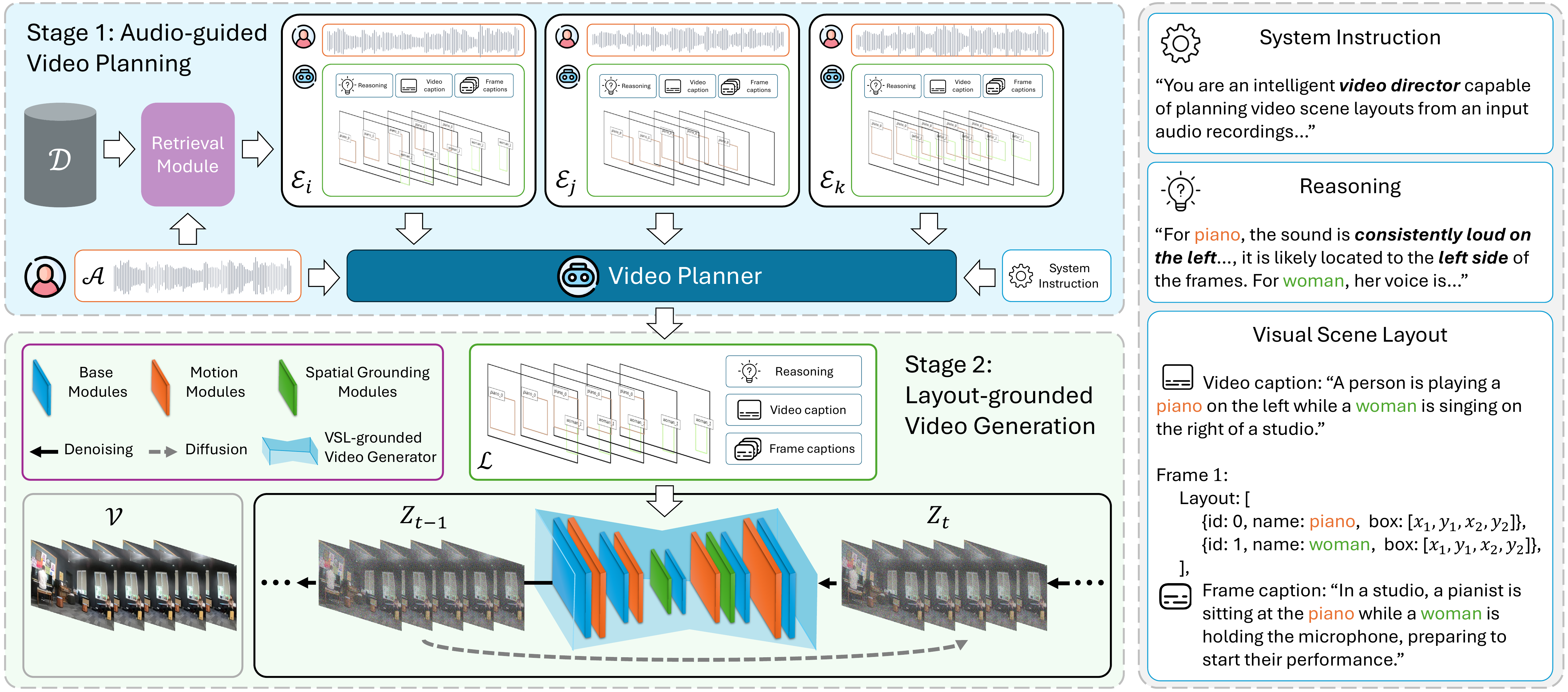}
  \caption{Illustration for the overall framework of SpA2V which is decomposed into two stages: Audio-guided Video Planning and Layout-grounded Video Generation. In the first stage (Section~\ref{sec:stage1}), given an input audio $\mathcal{A}$, we retrieve $k$ example conversations $\{\mathcal{E}_1, \mathcal{E}_2, \dots \mathcal{E}_k\}$ from candidate database $\mathcal{D}$ via Retrieval Module and feed them together with a System Instruction and the audio itself into the MLLM Video Planner to perform reasoning and generate a desired VSL sequence $\mathcal{L}$ containing $N$ consecutive keyframe layouts $\{\mathcal{L}_1, \mathcal{L}_2, \dots \mathcal{L}_N\}$ with respective global video caption and local frame captions. In the second stage (Section~\ref{sec:stage2}), the obtained VSL $\mathcal{L}$ and its captions are incorporated to guide a video diffusion model consisting of pretrained Base, Motion, and Spatial Grounding Modules to generate the final video $\mathcal{V}$ that is semantically and spatially coherent with the input $\mathcal{A}$.}
  \Description{}
  \label{fig:pipeline}
\end{figure*}
\section{Method}
 Although audio contains a significant amount of semantic and spatial information, effectively extracting and incorporating them for video generation are non-trivial and underexplored tasks. In this section, we describe how each stage of our proposed SpA2V framework tackles these challenges respectively. 
\subsection{Stage 1: Audio-guided Video Planning}
\label{sec:stage1}
\noindent\textbf{Overview.} In this stage, we introduce a novel task: generating video scene layouts (VSLs) depicting spatial arrangements of sounding objects presented in corresponding audio recordings. This task necessitates a model to first identify the categories of sounding sources (semantic component) and their respective locations and movements (spatial components) from the input audio. Then, the model must use this information to organize the objects into a coherent VSL, accurately reflecting their spatial correspondence with the audio and maintaining content consistency across the video sequence. Given these requirements, Multimodal Large Language Models (MLLMs) are particularly well-suited due to their strong multimodal understanding, reasoning abilities, and broad foundational knowledge. Consequently, we empirically investigate the potential of MLLMs to effectively address this challenging task. 

\noindent\textbf{Instruction Setup.} To generate an audio-conditioned VSL using an MLLM, we query it with a prompt consisting of three components: a system instruction, a set of example conversations, and a user-specified audio recording. The system instruction includes task definition and guidance regarding the desired behavior and response for each request that the MLLM must follow. Specifically, we instruct the MLLM to act as a \textit{video director} to plan VSLs that capture the content of the input audio recordings. We then outline the task requirements for the MLLM to fulfill, such as the expected layout format, coordinate system, canvas size, and number of frames. In complement, the example conversations provide the MLLM with reference query-response pairs, allowing it to efficiently learn and adapt to the given task. Finally, after supplying the above contextual information, we query the MLLM to perform completion on the user's input audio recording to generate the desired VSL.

\noindent\textbf{VSL Structure.} We ask the MLLM to generate VSLs according to a predefined template which is a connected sequence $\mathcal{L}$ of $N$ consecutive keyframe layouts $\{\mathcal{L}_1, \mathcal{L}_2, \dots, \mathcal{L}_N\}$. Every layout $\mathcal{L}_i$ contains a set of $N_i$ bounding boxes $\{\mathcal{B}_1, \mathcal{B}_2, \dots, \mathcal{B}_{N_i}\}$, each denotes a sounding object that occurs in the input audio. Each bounding box is represented by its location and size in numerical coordinates along with a labeling phrase that specifies the enclosed object. In addition, each box is assigned a unique numerical identifier, establishing and maintaining object correspondence across frames without the need for a dedicated box tracker. Finally, each VSL also entails a shared global video caption and a local frame caption for each keyframe describing the global content and local dynamic transition of the intended video creation. Note that in these captions, the MLLM has the freedom to bring about special information that cannot be inferred from input audio but is beneficial for video generation later such as visual appearance of sounding objects.

\noindent\textbf{Spatial Reasoning.} Spatial information can be inferred by reasoning on the fundamental spatial auditory cues, such as Interaural Time Difference (ITD), Interaural Level Difference (ILD), pitch and volume, and directional shift. ITD and ILD are typically used to infer the location of sounding objects, while pitch and volume often indicate their distance, and directional shift can imply their movement. To accurately deduce the corresponding spatial attributes and minimize spurious hallucinations, we explicitly instruct the MLLM in the system instruction to focus on analyzing these key indicators. Consequently, we ask the MLLM to output a brief statement summarizing its reasoning and the extracted spatial cues before generating the VSL to enhance the interpretability of the final response.  

\begin{figure}[t]
\includegraphics[width=\linewidth]{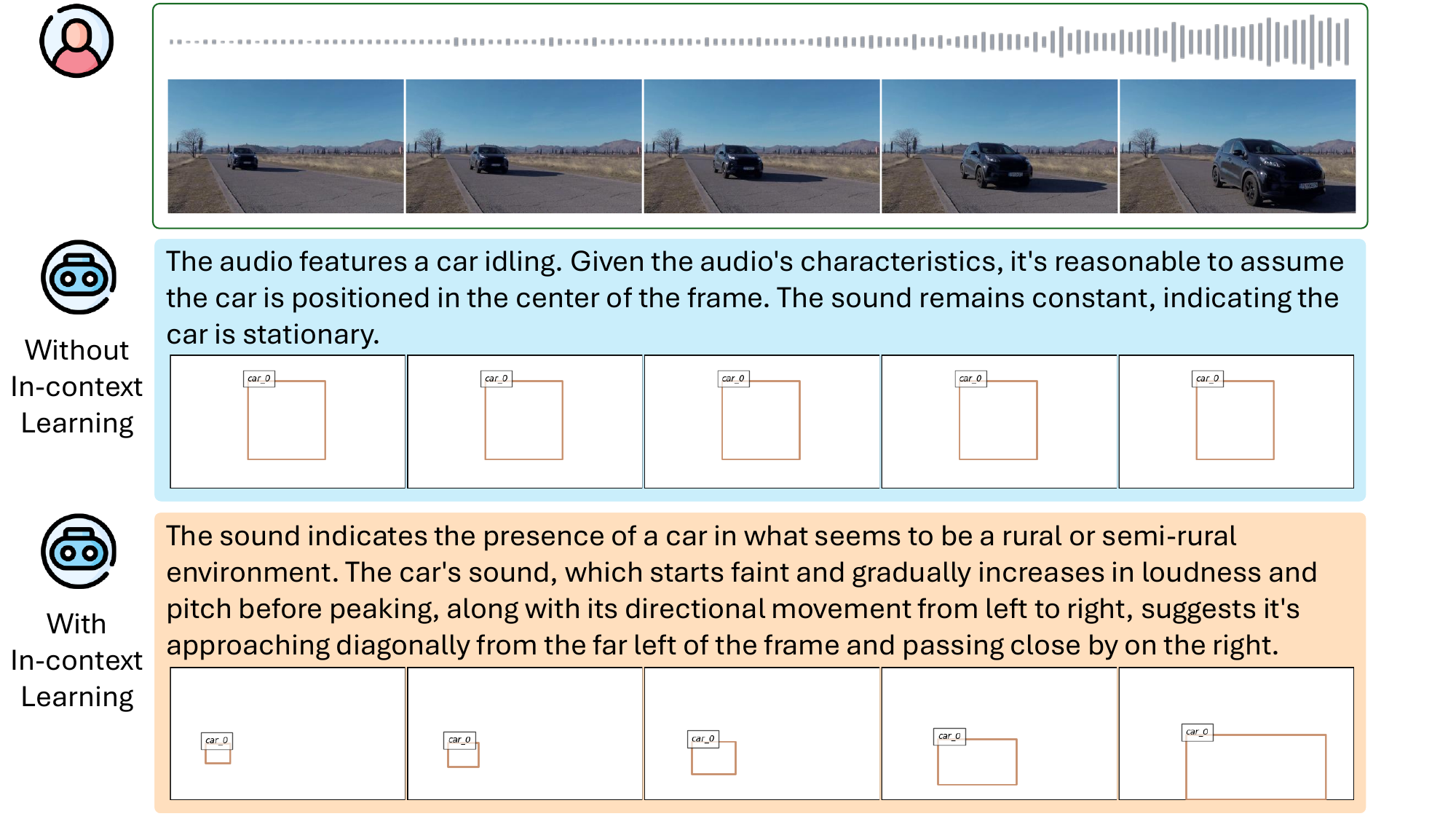}
  \caption{In-context Learning helps guide the MLLM to derive the correct spatial cues from the right physical sound properties and hence generate a highly-aligned VSL.}
  \Description{}
  \label{fig:incontext}
\end{figure}

\noindent\textbf{In-context Learning.} Solely relying on the system instruction to provide task descriptions and reasoning guidance may fall short in allowing the MLLM to comprehend our need for precise real-world understanding and reasoning on the aforementioned physical sound properties, causing it to hallucinate incorrect spatial information with fuzzy or non-sensical reasoning, and eventually generate VSLs misaligned with the input audio recordings as shown in Fig.~\ref{fig:incontext}. Inspired by~\cite{yang2024incontext, dong-etal-2024-survey, liu-etal-2022-makes, incontext} which show that In-context Learning can enhance the LLMs' task adaptability and compliance in various contexts, we employ it to further guide the behavior of the MLLM and mitigate mentioned problem. For each query, we provide the MLLM with example conversations, each including a reference prompt and a high-quality VSL with corresponding reasoning statement. Akin to~\cite{liu-etal-2022-makes}, we hypothesize that the more semantically similar the audio recordings of the examples to that of the query, the more informative it can be for the MLLM. Therefore, we conduct \textit{Top-k} Nearest Neighbor ($k$NN) search on CLAP~\cite{laionclap2023} embedding space in our Retrieval Module to select $k$ examples for each query.    
\subsection{Stage 2: Layout-grounded Video Generation}
\label{sec:stage2}
\noindent\textbf{Overview.} Leveraging the ability of MLLMs to generate semantically and spatially aligned VSLs and descriptive captions from auditory cues, we subsequently introduce an approach for video synthesis controlled by these VSLs in this stage. Our VSL-grounded Video Generator connects off-the-shelf layout-to-image and text-to-video diffusion models into a single pipeline. By combining their respective grounding and temporal modeling capabilities, our generator produces videos that adhere to the conditioned VSLs and entailed captions, thereby maintaining consistency with the input audio. Our method operates in a training-free manner that efficiently reduces computational cost and time, eliminates the need for extensive data annotation, and avoids potential catastrophic forgetting incurred by training. 

\noindent\textbf{Base Diffusion Model.} We build our VSL-grounded Video Generator based on the pre-trained text-to-image LDM~\cite{rombach2022high}, \textit{a.k.a} Stable Diffusion, of which the diffusion procedure follows the standard formulation
in~\cite{DDPM, song2021scorebased, sohl} that comprises a forward diffusion and a backward denoising process. Given a data sample $\mathbf{X} \sim \mathcal{P}(\mathbf{X})$, an autoencoder consisting of an encoder $\mathscr{E}$ and a decoder $\mathscr{D}$ will first project its latent correspondence $\mathbf{Z}_0=\mathscr{E}(\mathbf{X})$. Subsequently, the diffusion and denoising processes are conducted in latent space. In one hand, the forward diffusion is essentially a fixed Markov process of $T$ timesteps that gradually perturbs $\mathbf{Z}_0$ to yield $\mathbf{Z}_t$ via: 
\begin{equation}
  \mathbf{Z}_t = \sqrt{\bar{\alpha}_t}\mathbf{Z}_0 + \sqrt{1 - \bar{\alpha}_t}\epsilon, \epsilon\sim\mathcal{N}(0, \mathbf{I}),
\end{equation}
for $t=1, 2,\dots, T$. Here $\bar{\alpha}_t$ is pre-defined parameter which determines the noise strength at each timestep $t$. Eventually, $\mathbf{Z}_0$ turns into $\mathbf{Z}_T$ that is indistinguishable from a Gaussian noise. On the other hand, the backward process leverages a denoising network $\epsilon_\theta$ with training objective of minimizing:
\begin{equation}
\mathbb{E}_{t,\mathbf{C},\mathbf{Z}_t,\epsilon\sim\mathcal{N}(0, \mathbf{I})}||\epsilon-\epsilon_{\theta}(\mathbf{Z}_t,t,\tau_\theta(\mathbf{C}))||^2_2,
\end{equation}
where $\mathbf{C}$ is the condition and $\tau_\theta$ represents its encoder, to iteratively denoise $\mathbf{Z}_t$. Once the denoising is finished and a final clean latent $\hat{\mathbf{Z}}_0$ is obtained, the generated sample can be decoded via $\hat{\mathbf{X}}=\mathscr{D}(\hat{\mathbf{Z}}_0)$. In Stable Diffusion, $\epsilon_\theta$ adopts UNet~\cite{unet} architecture comprising of down/middle/up blocks each consisting of ResNet~\cite{resnet}, spatial self-attention layers, and cross-attention layers that incorporate text conditions. For conciseness, we call these blocks Base Modules in our VSL-grounded Video Generator, responsible for preserving pre-learned knowledge to generate diverse and high-fidelity samples guided by text prompts in image domain. 

\noindent\textbf{Integrating Grounding and Temporal Modeling.} With Stable Diffusion as the base model, we respectively integrate pretrained Temporal Modules and Grounding Modules from AnimateDiff~\cite{animatediff} and MIGC~\cite{migc} into our VSL-grounded Video Generator, enabling spatial grounding and motion modeling capabilities to synthesize high-quality videos aligned with input VSLs. Specifically, AnimateDiff proposes learning meaningful motion priors by injecting temporal transformer blocks, namely Motion Modules, to inflate Stable Diffusion, allowing it to generate motion dynamics of visual content over time while alleviating quality degradation. Meanwhile, MIGC incorporates a set of articulated instance enhancement attention layers, which we call Spatial Grounding Modules, into Stable Diffusion to enable precise generations of multiple instances in the resulting image following a layout input. Since only these external modules are trained to learn their designated objectives while the same base modules are kept frozen, we hypothesize then empirically verify that directly combining them into a single end-to-end pipeline, \textit{i.e.} our VSL-grounded Video Generator, can achieve both spatial grounding and motion modeling abilities.     
\begin{figure*}
  \includegraphics[width=\textwidth]{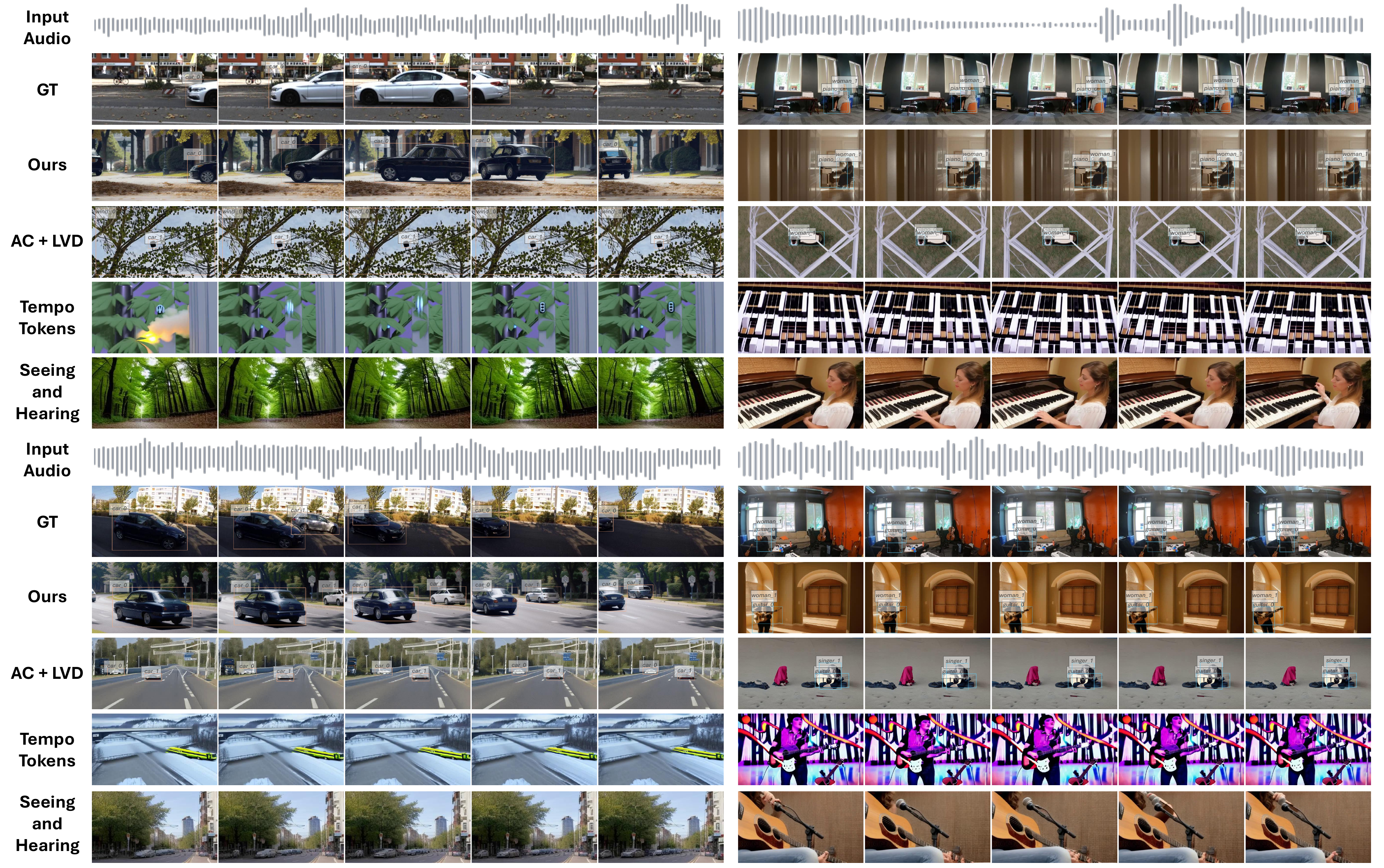}
  \caption{Qualitative comparisons of our SpA2V with prior SOTA works in audio-to-video generation. Here GT denotes ground-truth videos and VSLs for illustration of visual elements present in input audios. Zoom-in for details.}
  \Description{}
  \label{fig:qualitative}
\end{figure*}

\noindent\textbf{Video Generation with VSL Guidance.} Every VSL $\mathcal{L}$ comprises a sequence of $N$ consecutive keyframe layouts, each containing a set of object bounding boxes, a shared global video caption, and a local frame caption. To control our VSL-grounded Video Generator to synthesize a video of $n$ frames, we first perform temporal-wise linear interpolation on the coordinates of the bounding boxes for each object to obtain a denser VSL with expanded length of $n$ layouts. Each layout will then serve as a grounding signal for a corresponding frame. Since the base diffusion model uses text prompt as global condition for generation, we also input the global video caption of the VSL to preserve its pre-trained generative capability and maintain global consistency across generated frames. In addition, for the $N$ keyframes, we use their local frame caption as alternative to global caption that empirically helps produce better frame transitions with more natural local dynamics.
\begin{table*}[t]
    \begin{center}
    \resizebox{\textwidth}{!}{  
        \renewcommand{\arraystretch}{1.05}
        \begin{tabular}{c c c c c c|@{}|c c c c c c c c c|c c c c c c c c c}
        \Xhline{2.0\arrayrulewidth}
         \multirow{3}{*}{Method}& \multirow{3}{*}{Combo} & \multirow{3}{*}{IL Setup} & \multirow{3}{*}{Eg. Sel.} & \multirow{3}{*}{(M)LLM} & \multirow{3}{*}{$\tau$} & \multicolumn{9}{c|}{Stationary}& \multicolumn{9}{c}{Translational} \\
         & & & & & &  \multicolumn{3}{c}{\textit{MaxIoU} $\uparrow$} & \multicolumn{3}{c}{\textit{LTSim} $\uparrow$}& \multicolumn{3}{c|}{\textit{DocSim} $\uparrow$}& \multicolumn{3}{c}{\textit{MaxIoU} $\uparrow$} & \multicolumn{3}{c}{\textit{LTSim} $\uparrow$} & \multicolumn{3}{c}{\textit{DocSim} $\uparrow$} \\
        & & & & & & S & M & C & S & M & C & S & M & C & S & M & C & S & M & C & S & M & C \\
        \Xhline{2.0\arrayrulewidth}
         \multicolumn{2}{c}{AC + LVD~\cite{lian2024llmgrounded}} & 3-shot & Default & GPT4 & 0.5 & 0.92 & 0.96 & 0.94 & 40.48 & 46.32 & 42.97 & 4.40 & 4.91 & 4.61 & 1.79 & 1.51 & 1.77 & 47.51 & 45.17 & 47.35 & 3.69 & 3.98 & 3.71 \\
         \Xhline{2.0\arrayrulewidth}
        \multicolumn{1}{c|}{\multirow{15}{*}{SpA2V}} & {\cellcolor{red!25}Full} & \multirow{2}{*}{3-shot} & \multirow{2}{*}{$k$NN} & \multirow{4}{*}{G2.0F} & \multirow{4}{*}{0.5}& {\cellcolor{blue!25}20.16} & {\cellcolor{blue!25}18.55} & {\cellcolor{blue!25}19.45} & {\cellcolor{blue!25}76.73} & {\cellcolor{blue!25}74.43} & {\cellcolor{blue!25}75.73} & {\cellcolor{blue!25}15.69} & {\cellcolor{blue!25}15.06} & {\cellcolor{blue!25}15.47} & {\cellcolor{blue!25}22.62} & {\cellcolor{blue!25}20.13} & {\cellcolor{blue!25}22.24} & {\cellcolor{blue!25}77.55} & {\cellcolor{blue!25}73.90} & {\cellcolor{blue!25}77.21} & {\cellcolor{blue!25}16.77} & {\cellcolor{blue!25}13.66} & {\cellcolor{blue!25}16.50} \\
        \multicolumn{1}{c|}{\multirow{15}{*}{(Ours)}} & {\cellcolor{red!25}w/o SR} & & & & & 14.57 & 8.75 & 12.03 & 74.90 & 69.24 & 72.41 & 14.39 & 13.10 & 13.90 & 17.10 & 15.43 & 16.87 & 75.09 & 72.27 & 74.88 & 17.03 & 12.99 & 16.74\\
        \multicolumn{1}{c|}{} & {\cellcolor{red!25}w/o IL} & \multirow{2}{*}{N/A} & \multirow{2}{*}{N/A} & & & 3.93 & 1.71 & 3.00 & 62.64 & 56.01 & 59.84 & 4.55 & 4.18 & 4.40 & 5.19 & 2.22 & 4.96 & 62.72 & 56.36 & 62.26 & 6.07 & 4.71 & 5.98\\
        \multicolumn{1}{c|}{} & {\cellcolor{red!25}Vanilla} & & & & & 4.63 & 1.77 & 3.42 & 66.56 & 56.37 & 62.23 & 5.47 & 4.61 & 5.10 & 6.58 & 2.58 & 6.32 & 67.37 & 60.52 & 66.91 & 6.05 & 4.05 & 5.93\\
        \hhline{|~|-----------------------|}
        \multicolumn{1}{c|}{} & \multirow{3}{*}{Full} & {\cellcolor{yellow!25}3-shot} & \multirow{3}{*}{$k$NN} & \multirow{3}{*}{G2.0F} & \multirow{3}{*}{0.5} & 20.16 & 18.55 & 19.45 & 76.73 & 74.43 & 75.73 & 15.69 & 15.06 & 15.47 & 22.62 & 20.13 & 22.24 & 77.55 & 73.90 & 77.21 & 16.77 & 13.66 & 16.50 \\
        \multicolumn{1}{c|}{} & & {\cellcolor{yellow!25}2-shot} & & & & 14.46 & 8.03 & 11.72 & 74.58 & 70.59 & 72.86 & 15.01 & 14.30 & 14.72 & 11.27 & 11.28 & 11.26 & 73.35 & 71.35 & 73.24 & 14.70 & 12.51 & 14.54 \\
        \multicolumn{1}{c|}{} & & {\cellcolor{yellow!25}1-shot} & & & & 10.18 & 5.30 & 8.02 & 71.17 & 66.32 & 69.12 & 10.39 & 11.03 & 10.65 & 7.86 & 7.08 & 7.80 & 70.40 & 66.61 & 70.14 & 10.44 & 9.65 & 10.39 \\
        \hhline{|~|-----------------------|}
        \multicolumn{1}{c|}{} & \multirow{2}{*}{Full} & \multirow{2}{*}{3-shot} & {\cellcolor{green!25}$k$NN} & \multirow{2}{*}{G2.0F} & \multirow{2}{*}{0.5} & 20.16 & 18.55 & 19.45 & 76.73 & 74.43 & 75.73 & 15.69 & 15.06 & 15.47 & 22.62 & 20.13 & 22.24 & 77.55 & 73.90 & 77.21 & 16.77 & 13.66 & 16.50 \\
        \multicolumn{1}{c|}{} & & & {\cellcolor{green!25}Random} & & & 4.47 & 2.28 & 3.58 & 62.07 & 56.33 & 59.57 & 6.86 & 7.34 & 7.13 & 8.01 & 4.46 & 7.71 & 71.43 & 67.03 & 71.10 & 9.11 & 8.18 & 9.03 \\
        \hhline{|~|-----------------------|}
        \multicolumn{1}{c|}{} & \multirow{3}{*}{Full} & \multirow{3}{*}{3-shot} & \multirow{3}{*}{$k$NN} & {\cellcolor{purple!25}G2.0F} & \multirow{3}{*}{0.5} & 20.16 & 18.55 & 19.45 & 76.73 & 74.43 & 75.73 & 15.69 & 15.06 & 15.47 & 22.62 & 20.13 & 22.24 & 77.55 & 73.90 & 77.21 & 16.77 & 13.66 & 16.50 \\
        \multicolumn{1}{c|}{} & & & & {\cellcolor{purple!25}G1.5F} & & 7.04 & 3.13 & 5.43 & 70.19 & 65.05 & 68.11 & 12.24 & 11.44 & 11.91 & 6.40 & 4.41 & 6.26 & 71.55 & 67.38 & 71.25 & 8.96 & 8.74 & 8.95 \\
        \multicolumn{1}{c|}{} & & & & {\cellcolor{purple!25}G4oM} & & 17.15 & 11.78 & 14.54 & 72.16 & 68.11 & 70.21 & 13.96 & 11.92 & 13.09 & 12.40 & 8.69 & 12.11 & 66.23 & 63.45 & 66.04 & 9.91 & 8.13 & 9.75 \\
        \hhline{|~|-----------------------|}
        \multicolumn{1}{c|}{} & \multirow{3}{*}{Full} & \multirow{3}{*}{3-shot} & \multirow{3}{*}{$k$NN} & \multirow{3}{*}{G2.0F} & {\cellcolor{gray!25}0.5} & 20.16 & 18.55 & 19.45 & 76.73 & 74.43 & 75.73 & 15.69 & 15.06 & 15.47 & 22.62 & 20.13 & 22.24 & 77.55 & 73.90 & 77.21 & 16.77 & 13.66 & 16.50 \\
        \multicolumn{1}{c|}{} & & & & & {\cellcolor{gray!25}1.0} & 16.54 & 16.87 & 16.42 & 75.73 & 73.92 & 74.83 & 14.59 & 14.47 & 14.49 & 18.45 & 16.51 & 18.24 & 76.11 & 72.57 & 75.85 & 14.89 & 12.64 & 14.70 \\
        \multicolumn{1}{c|}{} & & & & & {\cellcolor{gray!25}1.5} & 14.09 & 14.31 & 14.22 & 74.76 & 73.04 & 74.02 & 13.76 & 13.63 & 13.68 & 15.87 & 16.74 & 15.84 & 75.39 & 72.19 & 75.11 & 14.05 & 12.33 & 13.91 \\
        \Xhline{2.0\arrayrulewidth}
        \end{tabular}
    }
    \end{center}
    \caption{Quantitative results and ablation analysis conducted for Audio-driven Video Planning in Stage 1. Here AC, SR, IL, and Eg. Sel. are shortened for Audio Captioning, Spatial Reasoning, In-context Learning, and Example Selection, respectively. $\tau$ denotes the temperature value of (M)LLM and $\uparrow$ indicates higher values are better. S and M represent subsets of data samples having single or multiple sounding sources, while C represents combinations of all scenarios. G2.0F, G1.5F, and G4oM stands for Gemini 2.0 Flash, Gemini 1.5 Flash, and GPT4o Mini accordingly.}
    \label{tab:stage1}
    \vspace{-1em}
\end{table*}
\section{Experiments}
\subsection{Setup}
\noindent\textbf{Benchmark.} 
As our proposed two-stage Audio $\rightarrow$ VSL $\rightarrow$ Video pipeline is novel, there is no existing benchmark suitable for evaluating our SpA2V framework. Therefore, we created AVLBench, a new benchmark specifically designed for our use case, curated from real-world stereo audio-video recording datasets~\cite{fairplay, vs13, urbansas, vehicle} spanning a variety of sound sources, including instruments and moving vehicles in indoor and outdoor environments. We begin by manually selecting recordings for which the audio contains strong semantic and spatial signals clearly indicating the sounding sources and their attributes within the video. After filtering, we apply flip and reverse augmentations with quality control to increase the data diversity while maintaining strong correspondence between auditory and visual elements. Subsequently, we use Track Anything~\cite{trackanything} to generate ground-truth VSLs by tracking the sounding objects in the videos. Since we need to provide SpA2V's Video Planner with example conversations for In-context Learning, we adopt LLaVA-OneVision~\cite{llavaonevision} to generate global video caption and local frame captions for each video. We also include an accurate manually written reasoning statement for every sample. Eventually, AVLBench contains 7274 testing samples, of which 4702 samples are used to assess scenarios of single or multiple instruments playing while having \textit{Stationary} motion in indoor settings, whereas the rest 2572 samples target cases of single or multiple vehicles with \textit{Translational} movement in outdoor settings.  

\noindent\textbf{Implementation Details.} The overall structure of our SpA2V framework is illustrated in Fig.~\ref{fig:pipeline}. In Stage 1, we select Gemini 2.0 Flash~\cite{gemini} as our MLLM Video Planner to balance cost-effectiveness and performance. For each input audio, we provide it with $k=3$ example conversations retrieved from the candidate database via Retrieval Module that performs a $k$NN search based on the similarity between CLAP embeddings of the input and the candidates. Subsequently, we prompt the Video Planner to generate VSL consisting of $N=5$ keyframe layouts of resolution $454\times256$ with a temperature of $\tau=0.5$ to control the randomness of its response. In Stage 2, Stable Diffusion 1.5~\cite{sd1.5} is adopted as the base diffusion model of our VSL-grounded Video Generator. We then follow default settings in MIGC~\cite{migc} and AnimateDiff~\cite{animatediff} to accordingly deploy Spatial Grounding Modules and Motion Modules onto the base model. With this complete architecture, our Video Generator performs inference to synthesize video of $n=16$ frames with resolution $512\times320$ conditioned on the VSL obtained from Stage 1. Unless otherwise specified, these settings are kept by default.

\noindent\textbf{Metrics.} In Stage 1, to measure the quality of the results in alignment with the input audio, we compute the similarity between the generated VSL and the ground-truth utilizing three metrics namely \textit{LTSim}~\cite{ltsim} , \textit{MaxIoU}~\cite{maxiou}, \textit{DocSim}~\cite{docsim}. These metrics are designed for image layouts that contain bounding boxes of close-set labels. To calculate the similarity between a pair of layouts $(\mathcal{L}, \mathcal{L}')$, they first match their enclosed set of bounding boxes $(\{\mathcal{B}_i\}, \{\mathcal{B}_j'\})$ then accumulate coordinate IoU scores of the matched boxes. Matching typically involves an indicator function $f_{abs}(\mathcal{B}_i, \mathcal{B}'_j) = \mathbb{I}_{\{c_i=c'_j\}}$ that fully ignores box pairs with different categories. Nevertheless, our method generates VSL which is essentially a sequence of image layouts consisting of bounding boxes with free labels. Therefore, we adjust this indicator function to a \textit{soft} version $f_{soft}(\mathcal{B}_i, \mathcal{B}'_j)=cosine(P(c_i), P(c'_j))$ that measures the similarity of the two categories $(c_i, c'_j)$ in the projector $P$'s embedding space~\cite{bge_m3}. We then follow the rest of calculations for all metrics and average the score across frames for each VSL. 

For Stage 2, we adopt the standard FVD~\cite{fvd} and AV-Align~\cite{tempotoken} to accordingly assess the overall content quality of generated videos and their temporal alignment with input audios. Especially, to evaluate spatial correspondence, we first utilize OV-AVSS~\cite{ovavss} to localize the input audios' sounding objects within the synthesized videos to obtain respective VSLs. Subsequently, we compute the \textit{LTSim}~\cite{ltsim} scores between these and the ground-truth VSLs.

\noindent\textbf{Baselines.} Since there is no previous work explicitly explores audio-driven video planning, we choose a relevant baseline named LVD~\cite{lian2024llmgrounded} for comparison in Stage 1. Note that this approach generates dynamic scene layout conditioning on text prompt, whereas our task requires audio as the sole input guidance. Therefore, we adopt an audio captioning (AC) model~\cite{gama} to generate a textual description for each input audio and feed them into LVD respectively. For Stage 2, we additionally compare our SpA2V framework with TempoTokens~\cite{tempotoken}, Seeing and Hearing~\cite{seeing-and-hearing}, and LTX~\cite{ltx}, alongside LVD for system-level evaluations of audio-to-video generation capabilities. TempoTokens follows the typical Audio $\rightarrow$ Video direct pipeline for generation. Meanwhile, since Seeing and Hearing and LTX use textual condition, we provide them with the same audio captions as LVD. Therefore, they can be categorized as two-stage Audio $\rightarrow$ Text $\rightarrow$ Video approaches, as shown in Fig.~\ref{fig:difference} respectively.
\begin{table*}[t]
    \begin{center}
    \resizebox{\textwidth}{!}{  
        \renewcommand{\arraystretch}{1.05}
        \begin{tabular}{c c c|c c c c c c c c c|c c c c c c c c c}
        \Xhline{2.0\arrayrulewidth}
         \multirow{3}{*}{Method}& \multirow{3}{*}{Cap. Sel.} & \multirow{3}{*}{VSL Sel.} & \multicolumn{9}{c|}{Stationary}& \multicolumn{9}{c}{Translational} \\
         & & &  \multicolumn{3}{c}{\textit{FVD} $\downarrow$} & \multicolumn{3}{c}{\textit{AV-Align} $\uparrow$}& \multicolumn{3}{c|}{\textit{LTSim} $\uparrow$}& \multicolumn{3}{c}{\textit{FVD} $\downarrow$} & \multicolumn{3}{c}{\textit{AV-Align} $\uparrow$} & \multicolumn{3}{c}{\textit{LTSim} $\uparrow$} \\
         & & & S & M & C & S & M & C & S & M & C & S & M & C & S & M & C & S & M & C \\
        \Xhline{2.0\arrayrulewidth}
         \multicolumn{3}{c|}{TempoTokens~\cite{tempotoken}} & 878.70 & 759.22 & 691.89 & 0.153 & 0.134 & 0.145 & 34.49 & 34.47 & 34.22 & 1549.51 & 1355.67 & 1462.94 & 0.179 & 0.171 & 0.179 & 29.61 & 28.90 & 29.56\\
         \multicolumn{3}{c|}{Seeing and Hearing~\cite{seeing-and-hearing}} & 715.77 & 708.58 & 664.87 & 0.111 & 0.105 & 0.109 & 36.47 & 41.81 & 38.72 & 1144.97 & 979.19 & 1049.42 & 0.151 & 0.127 & 0.149 & 29.45 & 33.88 & 29.76\\
         \multicolumn{3}{c|}{AC + LTX~\cite{ltx}} & 619.97 & 525.14 & 543.81 & 0.091 & 0.088 & 0.090 & 34.75 & 39.55 & 36.79 & 1094.19 & 1154.74 & 1022.49 & 0.156 & 0.130 & 0.154 & 32.57 & 37.61 & 32.92 \\
         \multicolumn{2}{c}{AC + LVD~\cite{lian2024llmgrounded}} & Gen. & 814.03 & 793.48 & 712.55 & 0.156 & 0.129 & 0.144 & 31.40 & 33.65 & 32.43 & 1306.68 & 793.48 & 1196.76 & 0.158 & 0.126 & 0.156 & 42.31 & 46.27 & 42.58 \\
         \Xhline{2.0\arrayrulewidth}
        \multicolumn{1}{c|}{\multirow{4}{*}{SpA2V}} & {\cellcolor{cyan!25}Mix} & \multirow{3}{*}{Gen.} & {\cellcolor{pink!25}776.63} & {\cellcolor{pink!25}527.31} & {\cellcolor{pink!25}633.05} & {\cellcolor{pink!25}0.186} & {\cellcolor{pink!25}0.155} & {\cellcolor{pink!25}0.173} & {\cellcolor{pink!25}46.22} & {\cellcolor{pink!25}50.62 } & {\cellcolor{pink!25}48.10} & {\cellcolor{pink!25}302.88} & {\cellcolor{pink!25}594.38} & {\cellcolor{pink!25}278.99} & {\cellcolor{pink!25}0.170} & {\cellcolor{pink!25}0.187} & {\cellcolor{pink!25}0.171} & {\cellcolor{pink!25}69.50} & {\cellcolor{pink!25}61.71} & {\cellcolor{pink!25}68.96}\\
        \multicolumn{1}{c|}{\multirow{4}{*}{(Ours)}} & {\cellcolor{cyan!25}Global} & & 779.08 & 529.57 & 637.84 & 0.184 & 0.153 & 0.170 & 45.07 & 49.90 & 47.12 & 308.44 & 596.19 & 282.39 & 0.170 & 0.178 & 0.171 & 69.49 & 59.48 & 68.79\\
        \multicolumn{1}{c|}{} & {\cellcolor{cyan!25}Local} & & 790.47 & 536.77 & 643.30 & 0.176 & 0.154 & 0.166 & 44.99 & 50.02 & 47.14 & 313.23 & 598.37 & 288.01 & 0.159 & 0.184 & 0.161 & 69.44 & 62.24 & 68.94\\
        \hhline{|~|--------------------|}
        \multicolumn{1}{c|}{} & \multirow{2}{*}{Mix} & {\cellcolor{brown!25}Gen.} & 776.63 & 527.31 & 633.05 & 0.186 & 0.155 & 0.173 & 46.22 & 50.62 & 48.10 & 302.88 & 594.38 & 278.99 & 0.170 & 0.187 & 0.171 & 69.50 & 61.71 & 68.96 \\
        \multicolumn{1}{c|}{} & & {\cellcolor{brown!25}GT} & 744.79 & 515.20 & 619.05 & 0.185 & 0.158 & 0.173 & 49.83 & 52.05 & 50.77 & 244.55 & 576.18 & 231.84 & 0.170 & 0.180 & 0.171& 78.67 & 65.13 & 77.72\\
        \Xhline{2.0\arrayrulewidth}
        \end{tabular}
    }
    \end{center}
    \caption{Quantitative results and ablation analysis conducted for Layout-grounded Video Generation in Stage 2 and system-wise comparisons. Here Cap. Sel. and VSL Sel. denotes Caption Selection and Video Scene Layout Selection, $\uparrow$ and $\downarrow$ indicate higher or lower values are better, and Gen. and GT are shortened for Generated and Ground-truth VSL. Besides, S and M represent subsets of data samples having single or multiple sounding sources, while C represents combinations of all scenarios.}
    \label{tab:stage2}
    \vspace{-1em}
\end{table*}
\subsection{Evaluation of Audio-guided Video Planning}
\noindent\colorbox{blue!25}{\textbf{Overall Results.}} As demonstrated in Tab.~\ref{tab:stage1} and Fig.~\ref{fig:qualitative}, our SpA2V framework can generate VSLs with high similarity to the ground-truth VSLs which indicate strong alignments to the input audios. SpA2V significantly outperforms the baseline of combining audio captioning with LVD~\cite{lian2024llmgrounded} in all metrics and test scenarios.   

\noindent\colorbox{red!25}{\textbf{Component Ablation.}} We ablate each component of the MLLM Video Planner to evaluate their effectiveness accordingly. As indicated in Tab.~\ref{tab:stage1}, both In-context Learning and Spatial Reasoning are crucial for the planner to appropriately adapt to the instructed task and generate high-quality VSLs, omitting either one will lead to significant performance degradations. Interestingly, Spatial Reasoning needs to be accompanied by In-context Learning to synergistically help the planner achieve the best performance. Incorporating it alone may detrimentally confuse the planner and lead to subpar performance compared to not integrating both (Vanilla). 

\noindent\colorbox{yellow!25}{\textbf{In-context Learning Setup.}} We assess the performance of the MLLM Video Planner when providing it with different numbers of example conversations. Compared to the zero-shot Vanilla, In-context Learning consistently brings improvements to the planner by delivering more context information via selective examples.

\noindent\colorbox{green!25}{\textbf{Example Selection.}} We aim to empirically verify our assumption in Section~\ref{sec:stage1} that the more reference audio recordings in the example conversations are semantically similar to that of the query, the better information it can bring to the MLLM Video Planner to generate higher quality VSLs. We replace the $k$NN Searching strategy in the Retrieval Module with a simple random selection while keeping other settings as default. As shown in Tab.~\ref{tab:stage1}, this adjustment severely harms the overall performance, highlighting the advantages of the $k$NN Searching strategy we adopt.

\noindent\colorbox{purple!25}{\textbf{Choices of MLLM.}} Since the design of our SpA2V is flexible, it allows better models selected as its components to attain better performance. Here we try to employ different state-of-the-art MLLMs as the Video Planner for SpA2V. Specifically, we conduct the same experiments but switch from the default Gemini 2.0 Flash to its predecessor Gemini 1.5 Flash and GPT4o Mini~\cite{gpt4o}. The results in Tab.~\ref{tab:stage1} demonstrate that the default option significantly exceeds these alternatives, making it the best choice to accomplish this task.

\noindent\colorbox{gray!25}{\textbf{Temperature.}} We test the performance of our SpA2V's MLLM Video Planner with different values for temperature which controls the randomness of its response with higher values being more creative while lower ones being more deterministic. Apparently, a low value of $0.5$ best suits our need for the task, as shown in Tab.~\ref{tab:stage1}.
\subsection{Evaluation of Layout-to-Video Generation}
\noindent\colorbox{pink!25}{\textbf{Overall Results.}} As shown in Tab.~\ref{sec:stage2} and Fig.~\ref{fig:qualitative}, our SpA2V framework can generate high-quality videos with compelling semantic and spatial correspondence to input audios across various scenarios. Meanwhile, the synthesized videos of prior works are prone to having limited semantic coherence and inconsistent spatial composition with input audios. Additionally, these methods tend to create videos with minimal dynamics and struggle with cases where sounding objects have large movements. These results highlight the superiority of our proposed SpA2V framework and its two-stage Audio $\rightarrow$ VSL $\rightarrow$ Video pipeline in harnessing informative auditory cues from input audios for video generation objectives. Besides, SpA2V also achieve competitive \textit{AV-Align} scores which imply strong temporal alignment between generated videos and input audios. We attribute this to the MLLM Video Planner which has the innate potential to capture temporal features in complement of semantic and spatial cues from input audios. These information will then be harmoniously organized into according VSLs and propagated to the subsequent video generation. 

\noindent\colorbox{cyan!25}{\textbf{Caption Selection.}} As indicated in Tab.~\ref{tab:stage2}, simultaneously utilizing shared global and local keyframe captions as text conditions alongside VSL is empirically effective in enhancing SpA2V's performance. While the former helps preserve the pre-trained generative capability of employed diffusion model and maintain global consistency, the latter encourages better frame transitions with more natural local dynamics across generated frame. 

\noindent\colorbox{brown!25}{\textbf{Impact of VSL Quality.}} To evaluate the importance of VSL quality in generating high-fidelity videos aligned with input audios, we skip the video planning steps in Stage 1 and directly use ground-truth VSLs as alternative control signals to guide the video synthesis process in Stage 2. As demonstrated in Tab.~\ref{tab:stage2}, such an adjustment substantially enhances overall performance, indicating that the better the quality of VSLs, the better results we can achieve. Since our two-stage pipeline is implementation-agnostic, this observation further implies that our SpA2V framework can continue to improve generation quality and audio-video alignment by adopting more capable MLLMs and video diffusion models in flexible manner.
\section{Conclusion}
We have presented \textbf{SpA2V}, the first framework capable of harnessing spatial auditory cues for audio-driven spatially-aware video synthesis. SpA2V decomposes the generation process into two stages: \textit{Audio-guided Video Planning} and \textit{Layout-grounded Video Generation}. In Stage 1, we adopt a SOTA MLLM as the Video Planner and instruct it to generate VSLs from input audios through a diligently designed prompting mechanism. In Stage 2, we propose an effective Video Generator which efficiently incorporates off-the-shelf diffusion models to synthesize videos grounded by the VSLs obtained from previous stage. The experimental results on our newly introduced AVLBench benchmark highlight the superiority of our SpA2V in producing videos with high semantic and spatial consistency to the input audios, outperforming previous methods by large margins. We hope that our pipeline will encourage further exploration into related areas of study in the future.

\noindent
\textbf{Acknowledgement.} This research was supported by the Innovation and Technology Fund of HKSAR (GHX/054/21GD), the Hong Kong SAR RGC Early Career Scheme (26208924), the National Natural Science Foundation of China Young Scholar Fund (62402408), and the HKUST Sports Science and Technology Research Grant (SSTRG24EG04). 

\bibliographystyle{ACM-Reference-Format}
\bibliography{acmart}

\appendix
\section{Additional Implementation Details}
\begin{figure}[t]
  \includegraphics[width=\linewidth]{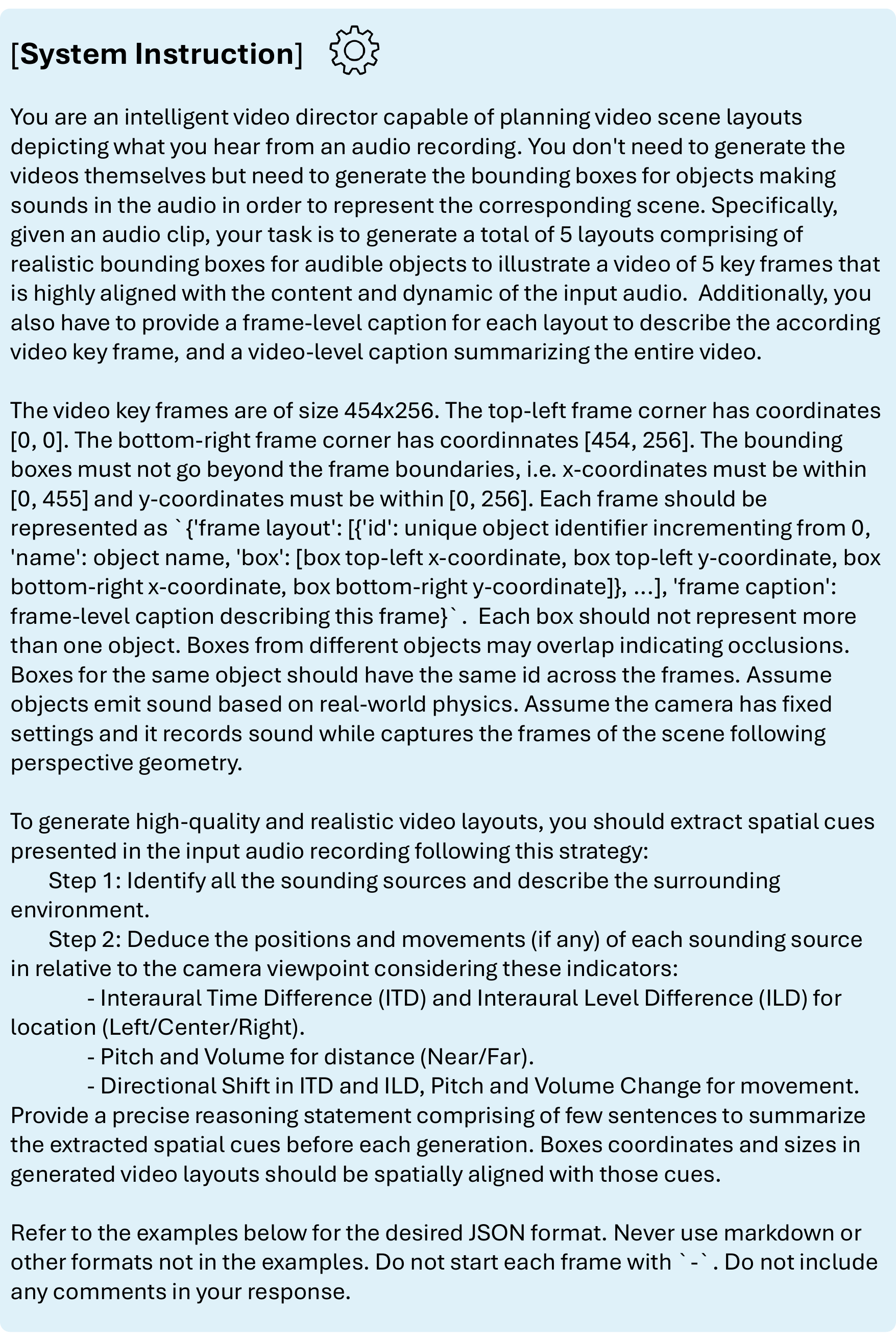}
  \caption{Our system instruction for the MLLM Video Planner to generate VSLs based on input audios.}
  \Description{}
  \label{fig:system_prompt}
\end{figure}

\noindent\textbf{System Instruction.} We present in Fig.~\ref{fig:system_prompt} the complete system instruction that we used to disclose task definition and guidelines to the MLLM Video Planner to control its behavior and response as we desired. This system instruction is inputted during the initialization of the MLLM.

\begin{figure}[t]
  \includegraphics[width=\linewidth]{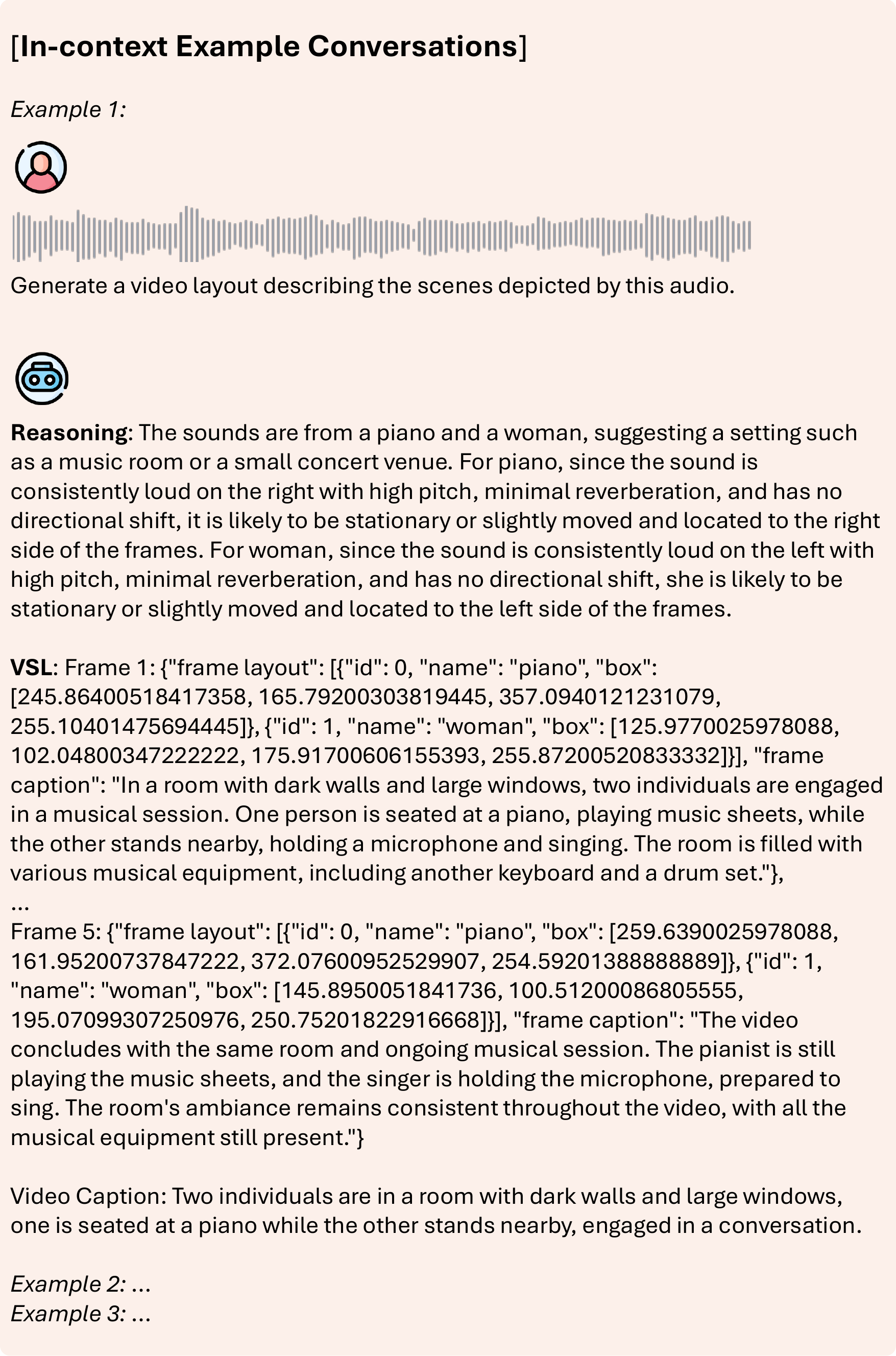}
  \caption{The template of our in-context example conversations to provide context for the MLLM Video Planner.}
  \Description{}
  \label{fig:incontext_template}
  \vspace{-0.5cm}
\end{figure}

\noindent\textbf{In-context Example Conversation.} In Fig.~\ref{fig:incontext_template}, we present the full template for each in-context example conversation that provides explicit context information for the MLLM to enhance its adaptability and adherence to the task. Each example contains a reference pair of user query and MLLM response comprised of a reasoning statement and the according visual scene layout (VSL). Every VSL in the examples follow the same structure as illustrated in Fig.~\ref{fig:incontext_template} and described in Section 3.1 in the main paper. 

\noindent\textbf{Motion and Spatial Grounding Modules.} Since our focus is to demonstrate the potential of our proposed Audio $\rightarrow$ Layout $\rightarrow$ Video direction for audio-driven video generation, we adopt the best configuration for the Motion and Spatial Grounding Modules from AnimateDiff and MIGC respectively for simplicity. Specifically, Motion Modules are inserted into every up- and down-sample block in Stable Diffusion's UNet, while Spatial Grounding Modules are deployed only on the middle block and the lowest-resolution up-sample block.
\section{Benchmark Construction}
\begin{figure}[t]
  \includegraphics[width=\linewidth]{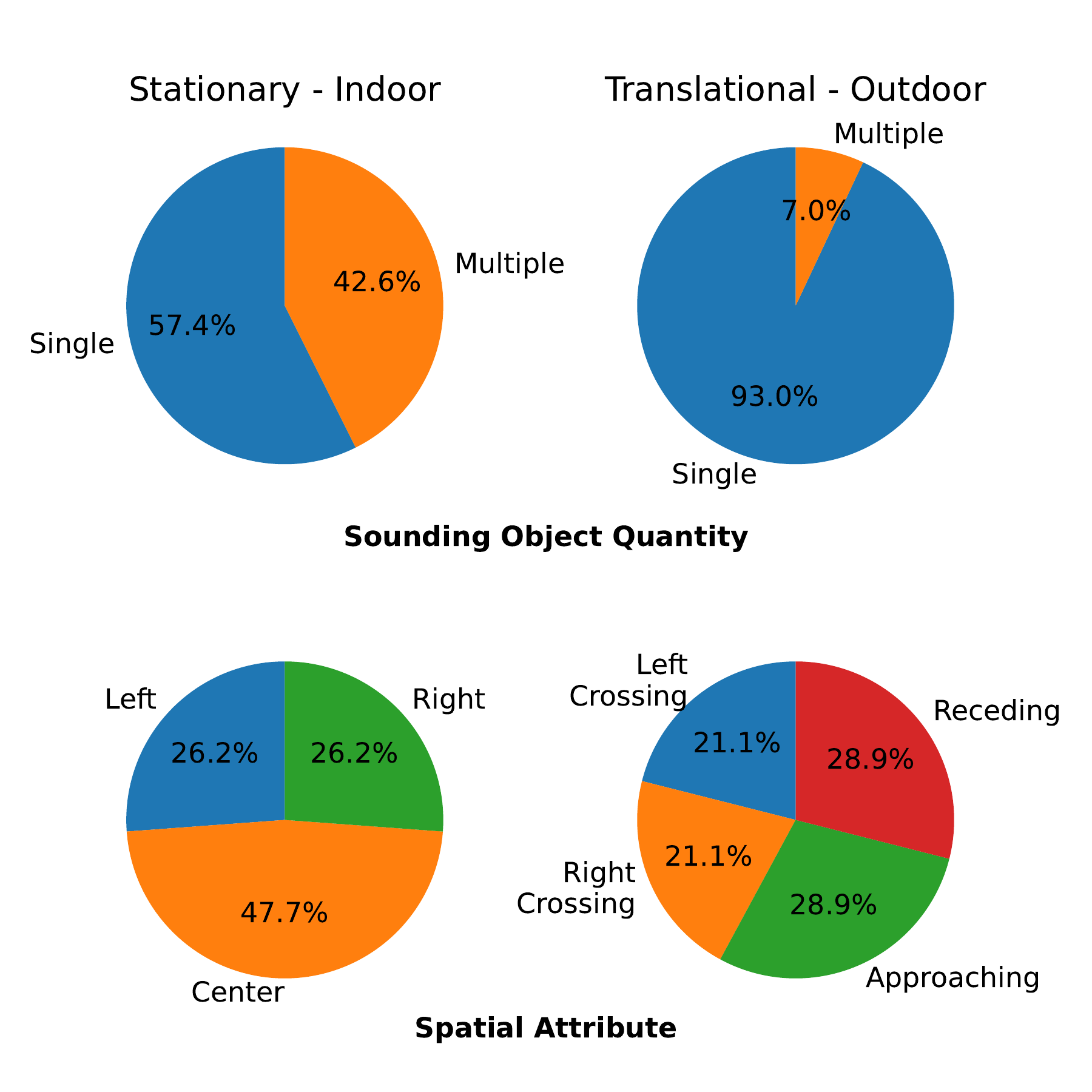}
  \caption{Breakdown statistics of AVLBench.}
  \Description{Description}
  \label{fig:distribution}
\end{figure}

Here we aim to provide more detailed information about the construction of our AVLBench benchmark specifically designed to assess Audio $\rightarrow$ VSL $\rightarrow$ Video generation abilities. Concretely, we build the benchmark following below four steps:
\begin{enumerate}
    \item \textbf{Sourcing.} We begin by curating data samples from existing datasets namely FAIR-Play~\cite{fairplay}, VS13~\cite{vs13}, Urbansas~\cite{urbansas}, and Freiburg Audio-Visual Vehicles~\cite{vehicle}. While the first contains various sample pairs of stereo audios and respective video recordings about instruments such as piano, trumpet, drums... being played in real-world indoor settings, the latter three target driving domains and their data samples capture moving vehicles in outdoor environments. We leverage these dataset for our use cases considering their high spatial alignment between auditory and visual elements of their sample pairs. 
    \item \textbf{Filtering.} We then manually select recordings and crop them into segments where there exists strong semantic and spatial signals in the audio that clearly indicates the sounding sources and their spatial attributes within the video, and remove noisy samples in which those signals are vague or unidentifiable.
    \item \textbf{Augmenting.} After careful filtering, we adopt flip and reverse augmentations with quality control to enrich the data diversity while preserving the strong correspondence between auditory and visual elements in the original samples. For flip, we apply it horizontally on the video frames while swapping the two channels of the paired audio for each sample. For reverse, we apply it on the temporal order of both video frames and audio. We observe that for audios containing sounds with high-frequencies such as instruments', applying reverse augmentation produce unnatural sounds with noisy artifacts, whereas it is not the case for low-level sound such as vehicle engines'. Therefore, we only apply flip augmentation for data samples originated from FAIR-Play, while we apply both augmentations for ones about moving vehicles.
    \item \textbf{Annotating.} Finally, we proceed to annotate each obtained sample to get their video scene layouts. Given the sounding sources in the audio, we use the Track Anything~\cite{trackanything} tool to track their locations and movements in the video. Besides, as required by the example conversations for In-context Learning of the MLLM Video Planner, we then utilize LLaVA-OneVision~\cite{llavaonevision} to generate global video caption and local frame captions on the video, and include a manually-written reasoning statement for every sample. 
\end{enumerate}
Eventually, AVLBench comprises 7274 testing samples, with 4702 samples designed to assess scenarios involving single or multiple instruments played in stationary indoor settings. The remaining 2572 samples focus on cases of single or multiple vehicles exhibiting translational movement in outdoor environments. The breakdown statistics of AVLBench on scene distribution and spatial attribute are detailed in Tab.~\ref{tab:breakdown_stats1},~\ref{tab:breakdown_stats2} and Fig.~\ref{fig:distribution}. Note that due to the noisy nature of outdoor environments, this domain mainly contains samples with single sounding vehicle after filtering. Besides, the spatial attribute statistics are accumulated per sounding object.
\section{Additional Experiments}
\noindent\textbf{Retrieval with more neighbors.} We conduct additional analysis on In-context Learning with more example conversations retrieved and provided to the MLLM Video Planner in Stage 1. The results shown in Tab~\ref{tab:add_knn} indicate that $k=3$ is the optimal setting, and further increasing the number of neighbors saturate the performance. Therefore, we use 3 in-context examples by default in the paper.

\noindent\textbf{Impact of retrieval database size and quality.} We conduct additional experiments under two adverse settings that reduce the size and diminish the quality of the retrieval database to demonstrate the influence of these factors on the planning stage. In the first setting, we halve the size of the retrieval database for each query randomly. In the second one, to exacerbate the challenge, we only use a fixed set of example conversations to provide in-context information for every query. The performance drops shown in Tab.~\ref{tab:half} indicate that the retrieval effectiveness is indeed sensitive against the database size and quality. This is reasonable because reducing the size and quality of retrieval corpora decreases the likelihood of retrieving relevant examples and creates more challenging out-of-domain scenarios. In future work, our aims are to expand the current dataset to cover more domains and scenarios as well as train an MLLM specialist for audio-driven video planning, mitigating this issue and enhancing the framework's feasibility in practice.

\noindent\textbf{User study.} To subjectively assess the performance of our SpA2V compared to other methods, we invite 25 users to participate in a user study. We ask each user to complete a set of 20 ranking questions, each composed of a query sample randomly selected from our benchmark and 5 videos generated by SpA2V and other 4 baselines. Users are required to rank them based on two criteria: (1) visual quality, and (2) audio-video alignment, with 1 indicating the best and 5 denoting the worst. The average ranking scores in Tab.~\ref{tab:userstudy} highlight the preference of users for the videos generated by our SpA2V over the others in both criteria.

\begin{table}[t]
    \begin{center}
    \resizebox{0.475\textwidth}{!}{  
        \renewcommand{\arraystretch}{1.05}
        \begin{tabular}{c c c|c c c|c}
        \Xhline{2.0\arrayrulewidth}
           \multicolumn{3}{c|}{Stationary}& \multicolumn{3}{c|}{Translational} & \multirow{2}{*}{Total} \\
          Single & Multiple & Subtotal &Single & Multiple & Subtotal & \\
          \Xhline{2.0\arrayrulewidth}
          2698 & 2004 & 4702 & 2392 & 180 & 2572 & 7274\\
        \Xhline{2.0\arrayrulewidth}
        \end{tabular}
    }
    \end{center}
    \caption{Statistics on scene distribution.}
    \label{tab:breakdown_stats1}
    \vspace{-0.5cm}
\end{table}
\begin{table}[t]
    \begin{center}
    \resizebox{0.475\textwidth}{!}{  
        \renewcommand{\arraystretch}{1.05}
        \begin{tabular}{c c c c|c c c c c|c}
        \Xhline{2.0\arrayrulewidth}
           \multicolumn{4}{c|}{Stationary}& \multicolumn{5}{c|}{Translational} & \multirow{3}{*}{Total} \\
          \multirow{2}{*}{Left} & \multirow{2}{*}{Center} & \multirow{2}{*}{Right} & \multirow{2}{*}{Subtotal} & Left & Right & \multirow{2}{*}{Approaching} & \multirow{2}{*}{Receding} & \multirow{2}{*}{Subtotal} \\
          & & & & Crossing & Crossing & & & & \\
          \Xhline{2.0\arrayrulewidth}
         3083 & 5616 & 3083 & 11782 & 582 & 582 & 800 & 800 & 2764 & 14546\\
        \Xhline{2.0\arrayrulewidth}
        \end{tabular}
    }
    \end{center}
    \caption{Statistics on spatial attribute.}
    \label{tab:breakdown_stats2}
    \vspace{-0.5cm}
\end{table}
\begin{table}[t]
    \begin{center}
    \resizebox{0.475\textwidth}{!}{  
        \renewcommand{\arraystretch}{1.05}
        \begin{tabular}{c|c c c|c c c}
        \Xhline{2.0\arrayrulewidth}
         \multirow{2}{*}{IL Setup}&  \multicolumn{3}{c|}{Stationary}& \multicolumn{3}{c}{Translational} \\
          &\textit{MaxIoU} $\uparrow$ & \textit{LTSim} $\uparrow$& \textit{DocSim} $\uparrow$&\textit{MaxIoU} $\uparrow$ & \textit{LTSim} $\uparrow$& \textit{DocSim} $\uparrow$ \\
          \Xhline{2.0\arrayrulewidth}
         0-shot& 3.00 &	59.84	& 4.40	&4.96&62.26	&5.98\\
         1-shot& 8.02 &	69.12	& 10.65	& 7.80	&70.14	&10.39\\
         2-shot& 11.72 &	72.86	& 14.72	&11.26 &	73.24	&14.54\\
         3-shot& 19.45 &	75.73	& 15.47	&22.24&77.21	&16.50\\
         5-shot& 16.77& 74.18&	15.14 &20.21 &	76.41	& 17.01\\
         7-shot& 16.49 & 74.29 & 15.24 &19.63 &	76.16	&16.93\\
        \Xhline{2.0\arrayrulewidth}
        \end{tabular}
    }
    \end{center}
    \caption{Planning results with different retrieval settings.}
    \label{tab:add_knn}
    \vspace{-0.5cm}
\end{table}
\begin{table}[t]
    \begin{center}
    \resizebox{0.475\textwidth}{!}{  
        \renewcommand{\arraystretch}{1.05}
        \begin{tabular}{c|c c c|c c c}
        \Xhline{2.0\arrayrulewidth}
         \multirow{1}{*}{Retrieval}&  \multicolumn{3}{c|}{Stationary}& \multicolumn{3}{c}{Translational} \\
          Setup&\textit{MaxIoU} $\uparrow$ & \textit{LTSim} $\uparrow$& \textit{DocSim} $\uparrow$&\textit{MaxIoU} $\uparrow$ & \textit{LTSim} $\uparrow$& \textit{DocSim} $\uparrow$ \\
          \Xhline{2.0\arrayrulewidth}
         Full Size & 19.45 &	75.73	& 15.47	&22.24&77.21	&16.50\\
         Half Size & 12.76 & 71.77 & 13.69 & 17.12 & 75.17 & 15.00\\
         Fixed Set & 3.68 & 59.44 & 7.90 & 7.58 & 71.59 & 9.14 \\
        \Xhline{2.0\arrayrulewidth}
        \end{tabular}
    }
    \end{center}
    \caption{Impact of retrieval database size and quality.}
    \label{tab:half}
    \vspace{-0.5cm}
\end{table}

\begin{table}[t]
    \begin{center}
    \resizebox{0.475\textwidth}{!}{  
        \renewcommand{\arraystretch}{1.05}
        \begin{tabular}{c|c c c c c}
        \Xhline{2.0\arrayrulewidth}
         \multirow{1}{*}{Method}&  \multicolumn{1}{c}{SpA2V}& \multicolumn{1}{c}{Seeing and Hearing}& \multicolumn{1}{c}{AC + LTX}& \multicolumn{1}{c}{AC + LVD}& \multicolumn{1}{c}{TempoTokens} \\
         \Xhline{2.0\arrayrulewidth}
          Visual quality $\downarrow$& 1.97 & 2.79 & 2.79 & 3.20 & 4.24 \\
          Audio-video alignment $\downarrow$ & 1.95 & 2.88 & 2.92 & 3.34 & 3.91\\
        \Xhline{2.0\arrayrulewidth}
        \end{tabular}
    }
    \end{center}
    \caption{User preference of SpA2V over prior works.}
    \label{tab:userstudy}
    \vspace{-0.5cm}
\end{table}
\begin{table}[t]
    \begin{center}
    \resizebox{0.475\textwidth}{!}{  
        \renewcommand{\arraystretch}{1.05}
        \begin{tabular}{c|c c c c c}
        \Xhline{2.0\arrayrulewidth}
         DeSync $\downarrow$&  \multicolumn{1}{c}{SpA2V}& \multicolumn{1}{c}{Seeing and Hearing}& \multicolumn{1}{c}{AC + LTX}& \multicolumn{1}{c}{AC + LVD}& \multicolumn{1}{c}{TempoTokens} \\
         \Xhline{2.0\arrayrulewidth}
          Stationary & 1.758 & 1.823 & 1.726 & 1.849 & 1.782 \\
          Translational & 1.136 & 1.584 & 1.658 & 1.620 & 1.782\\
        \Xhline{2.0\arrayrulewidth}
        \end{tabular}
    }
    \end{center}
    \caption{Additional quantitative results.}
    \label{tab:desync}
    \vspace{-0.5cm}
\end{table}
\noindent\textbf{Extra quantitative evaluation.} Since AV-Align~\cite{tempotoken} is known to not work well in complex scenes, we additionally use DeSync metric which leverages Syncformer~\cite{syncformer} to measure audio-video temporal misalignment and show the
results in Tab.~\ref{tab:desync}. As consistently observed, our SpA2V achieves
competitive performance that indicates strong temporal alignment
between the generated videos and input audios. 

\noindent\textbf{Ablation on Motion and Spatial Grounding Modules.} Since removing the Motion Modules will degrade our Layout-to-Video generator into a Layout-to-Image generator that deviates from our video synthesis objective, we omit ablating these modules. We only conduct additional analysis to ablate Spatial Grounding Modules which will degrade our generator into a Text-to-Video model. The results shown in Tab.~\ref{tab:no_layout} highlight the importance of these modules in achieving better video generation quality and especially semantic and spatial alignment with input audios.  
\section{Limitations and Future Work}
\begin{figure}[t]
  \includegraphics[width=\linewidth]{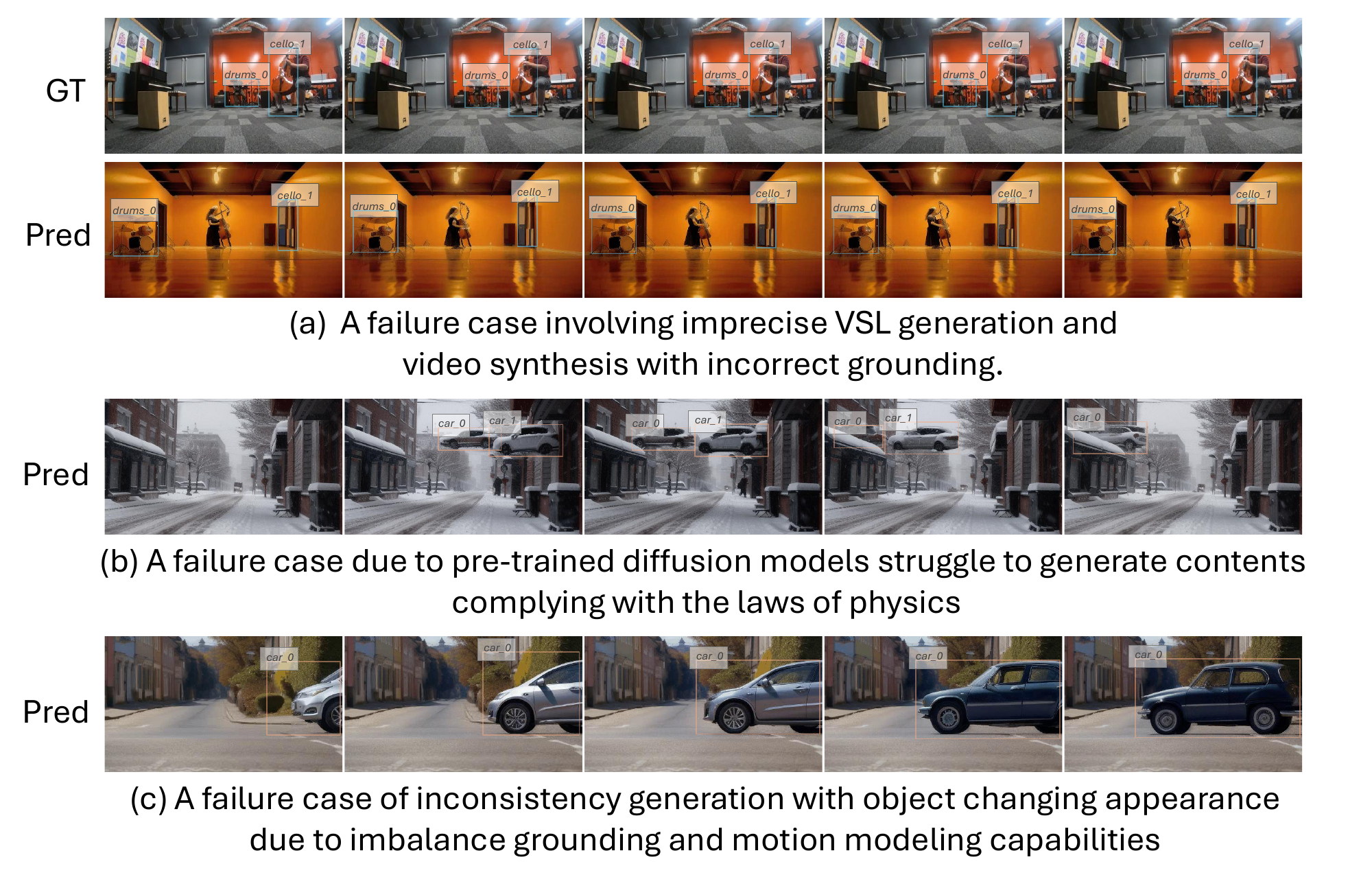}
  \caption{Illustration for limitations of our SpA2V.}
  \Description{Description}
  \label{fig:limitation}
\end{figure}

Although SpA2V introduces a novel two-stage Audio $\rightarrow$ VSL $\rightarrow$ Video pipeline for semantically and spatially aligned audio-driven video generation and achieve promising results that outperforms prior methods, there is still much room for further improvements. Firstly, as SpA2V involves two stages, failures in either stage will be detrimental to the whole generation process. For example, an incorrect VSL generated by the Video Planner in Stage 1 will inevitably lead to a synthesized video with misalignment in Stage 2 as shown in Fig.~\ref{fig:limitation}~(a). Secondly, since our SpA2V framework adopts pre-trained MLLMs and diffusion models as its Video Planner and Video Generator, it also inherits their existing limitations and its performance is hence heavily reliant on them. If they struggle to respond properly to a specific conditional guidance and fall short to generate accurate contents, such issue is likely to be propagated to SpA2V as shown in Fig.~\ref{fig:limitation}~(b). We anticipate that these two challenges can be appropriately mitigated by adopting or introducing more powerful models as the components of SpA2V. Finally, since we directly incorporate Spatial Grounding and Motion Modules from MIGC~\cite{migc} and AnimateDiff~\cite{animatediff} although they are trained on datasets, such domain gap can lead to the imbalance between grounding and motion modeling capabilities of the Video Generator, causing it to produce videos with inconsistency problems like having objects changing appearance over time as shown in Fig.~\ref{fig:limitation}~(c). We contemplate that a further finetuning step for the whole framework using techniques such as LoRA~\cite{hu2022lora} can help alleviate this issue and leave this exploration for future research.
\begin{table}[t]
    \begin{center}
    \resizebox{0.475\textwidth}{!}{  
        \renewcommand{\arraystretch}{1.05}
        \begin{tabular}{c|c c c c|c c c c}
        \Xhline{2.0\arrayrulewidth}
         \multirow{2}{*}{Method}&  \multicolumn{4}{c|}{Stationary}& \multicolumn{4}{c}{Translational} \\
          &\textit{FVD} $\downarrow$ & \textit{AV-Align} $\uparrow$&\textit{LTSim} $\uparrow$& \textit{DeSync} $\downarrow$&\textit{FVD} $\downarrow$ & \textit{AV-Align} $\uparrow$&\textit{LTSim} $\uparrow$& \textit{DeSync} $\downarrow$ \\
          \Xhline{2.0\arrayrulewidth}
         Full& 633.05 &	0.173	& 48.10	&1.758 &	278.99	&0.171	&68.96	& 1.136\\
         No Spatial Grounding & 730.26 &	0.063 & 42.02 & 1.773 & 760.02 & 0.121 & 63.89& 1.486\\
        \Xhline{2.0\arrayrulewidth}
        \end{tabular}
    }
    \end{center}
    \caption{Ablation on Spatial Grounding Modules.}
    \label{tab:no_layout}
    \vspace{-0.5cm}
\end{table} 
\section{Societal Impacts}
SpA2V empowers individuals, regardless of their video-photography ability, to generate videos that are both semantically and spatially aligned with audio inputs. However, employing our framework carries potential risks. It could be misused for malicious purposes, such as inappropriate content creation or the dissemination of misinformation. Furthermore, given our reliance on pre-trained MLLMs and diffusion models, our framework may inherit biases present in their training data, potentially perpetuating harmful stereotypes. While generated content may currently be readily distinguishable from original works, future technological advancements may blur this distinction, making infringement more difficult to detect. Therefore, we strongly urge users to exercise caution and utilize this method only for legitimate purposes.

\end{document}